\documentclass{vgtc}                          

\graphicspath{{figures/}{pictures/}{images/}{./}} 

\usepackage{times}                     

\usepackage{tabu}                      
\usepackage{booktabs}                  
\usepackage{lipsum}                    
\usepackage{mwe}                       

\usepackage{mathptmx}                  

\usepackage{amsmath}
\usepackage{multirow}
\usepackage{listings}
\usepackage{xcolor}
\usepackage{xspace}
\usepackage{tikz}
\usepackage{enumitem}
\usepackage{makecell}
\usepackage{tabularx}
\usepackage{subcaption}
\usepackage{fancyvrb}
\usetikzlibrary{shadings}
\usetikzlibrary{shapes}

\onlineid{8774}

\vgtccategory{Research}

\vgtcinsertpkg

\newcommand{\sys}{\textsc{ChartRevive}\xspace}

\definecolor{cback}{HTML}{E9ECEF}
\definecolor{cframe}{HTML}{495057}
\definecolor{clightblue}{HTML}{DDEFFC}
\definecolor{clightgreen}{HTML}{DFFFE5}
\definecolor{clightyellow}{HTML}{FFF4DB}

\definecolor{cback1}{HTML}{F8F9FA} 
\definecolor{cframe1}{HTML}{ADB5BD} 

\definecolor{cback2}{HTML}{F1F3F5} 
\definecolor{cframe2}{HTML}{CED4DA} 

\definecolor{cback3}{HTML}{EBF5FB} 
\definecolor{cframe3}{HTML}{A9CCE3} 

\definecolor{cback4}{HTML}{FDF3E7} 
\definecolor{cframe4}{HTML}{F5CBA7} 

\newcommand*\circled[1]{\tikz[baseline=(char.base)]{
    \node[shape=rectangle,rounded corners=1.5pt,fill=cback,text=black,draw=cframe,inner sep=1pt] (char) {#1};}}

\newcommand*\circledVarD[1]{\tikz[baseline=(char.base)]{
    \node[shape=rectangle,rounded corners=1.5pt,fill=cback4,text=black,draw=cframe4,inner sep=1pt] (char) {#1};}}

\title{ChartRevive: Reconstructing Data Visualizations from Chart Images Using MLLMs}

\author{Yuki Ueno\thanks{e-mail: yueno@asu.edu}\\ %
        \scriptsize Arizona State University %
\and Aditeya Pandey\thanks{e-mail: aditeya.pandey@lilly.com}\\ %
     \scriptsize Eli Lilly and Company %
}

\teaser{
  \centering
  \includegraphics[width=\linewidth]{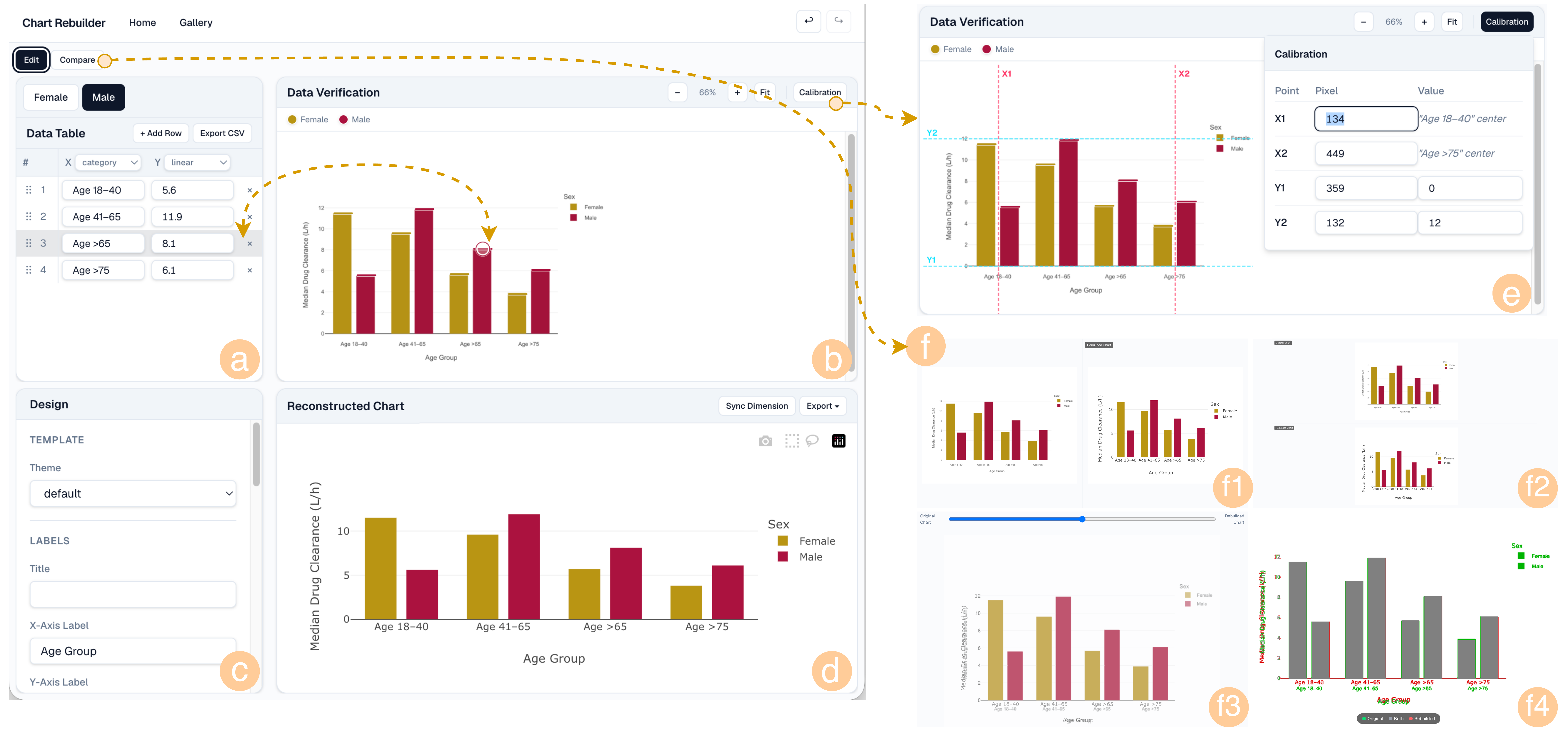}
  \caption{\sys supports chart reconstruction through automatic data and design extraction combined with an
  interactive verification interface comprising: (a) a Data Table panel for displaying and editing extracted data; (b) a Data Verification panel for verifying and correcting extracted data; (c) a Design panel for editing chart design; (d) a Reconstructed Chart panel displaying the Plotly-rendered chart; (e) a Calibration Tool for aligning the extracted chart with the original image; and (f) a Comparison View offering four comparison modes: (f1) Horizontal Split, (f2) Vertical Split, (f3) Fade, and (f4) Highlight.}
  \label{fig:teaser}
}

\abstract{Static chart images are widely used in scientific publications, business reports, and presentations, yet recovering both the underlying data and visual design from chart images remains a labor-intensive manual process, making them difficult to reuse. While prior work has primarily focused on data extraction, the extraction of visual design specifications, including colors, marker shapes, and axis configurations, remains underexplored. To identify a suitable model for chart reconstruction, we systematically benchmark five multimodal large language models (MLLMs) across five basic chart types on both data and design extraction tasks. Our evaluation shows that textual and categorical information can generally be extracted reliably, whereas numeric and spatial information remain challenging. Among the evaluated models, GPT-5.4 achieves the best overall performance and is adopted as the backbone of our system. Guided by these findings, we present \sys, a mixed-initiative system that combines MLLM-based extraction with an interactive verification interface, supporting users to efficiently inspect, correct, and refine reconstructed charts through overlay-based verification and real-time rebuilding.} 


\begin{document}

\emergencystretch=1em


\maketitle

\section{Introduction}

Static chart images are ubiquitous in scientific publications, business reports, and presentations~\cite{huan_pixels_2025, islam_overview_2019}. 
However, without underlying data tables, reusing and adapting them remains challenging~\cite{web_plot_digitizer, chai_crowdchart_2021, luo_chartocr_2021}. 
To analyze the underlying data or apply a consistent visual style across multiple figures (e.g., following company brand guidelines), users must first extract both the underlying data and the chart's visual design (e.g., colors, marker shapes, line styles, and axis configurations) from the image. 
This reverse-engineering process is typically performed manually and is both labor-intensive and time-consuming.

While prior research has primarily focused on extracting data from chart images~\cite{web_plot_digitizer, he_making_2026, liu_deplot_2023, masson_chartdetective_2023}, comparatively little attention has been paid to extracting their visual design. 
We observe that design extraction addresses practical needs that data extraction alone cannot fulfill. 
For example, when recreating a chart with updated data, preserving its original visual design enables consistent styling across a collection of figures. 
Likewise, maintaining visual fidelity is essential for verifying the correctness of reconstructed charts.

Recent advances in multimodal large language models (MLLMs) offer a promising opportunity to automate chart extraction. 
However, our formative study with professional designers revealed three major limitations of purely MLLM-based approaches: the extracted results are difficult to verify, iterative refinement is cumbersome, and the overall workflow is fragmented across multiple disconnected steps.
Motivated by these findings, we propose a mixed-initiative approach that combines MLLM-based extraction with an interactive verification interface, enabling users to efficiently inspect, correct, and refine the extracted results.

To select an appropriate extraction engine for our system, we benchmarked five commercial MLLMs on both data and design extraction tasks across five basic chart types. 
Our evaluation showed that textual and categorical information can generally be extracted reliably, whereas numeric and spatial information remains challenging for both data and design extraction. 
Among the evaluated models, \texttt{GPT-5.4} consistently achieved the best overall performance and was therefore adopted in our system.

The main contributions of this paper are as follows:

\begin{itemize}[itemsep=2pt]
    \item A formative study revealing the limitations of MLLM-based chart extraction from the perspective of chart designers.
    \item A benchmark of five commercial MLLMs for extracting chart data and visual design across five basic chart types.
    \item A mixed-initiative system that integrates MLLM-based extraction with an interactive verification interface to support efficient and quick reconstruction.
\end{itemize}
\section{Related Work}

Research on chart image extraction broadly falls into two categories: \textit{data extraction}, which recovers the underlying values, and \textit{design specification extraction}, which recovers visual properties such as colors, marker shapes, line styles, and axis configurations.

Early approaches to chart data extraction relied on OCR combined with chart-type-specific heuristics. Interactive tools such as WebPlotDigitizer~\cite{web_plot_digitizer} provide high accuracy but require tedious point-by-point annotation, while semi-automatic approaches such as ChartDetective~\cite{masson_chartdetective_2023} operate only on vector graphics (e.g., SVG), limiting their applicability to published raster figures. More recently, chart-to-table extraction has emerged as a standardized benchmark task. DePlot~\cite{liu_deplot_2023} decomposes chart reasoning into two stages: converting a chart image into a linearized table and reasoning over the table with LLMs, achieving strong performance on chart question answering while assuming that data labels are visible. He et al.~\cite{he_making_2026} systematically evaluated MLLMs for chart data extraction and showed that current models struggle with numerical accuracy and visual grounding, highlighting the need for mixed-initiative workflows with human verification.

Compared with data extraction, design specification extraction has received far less attention, despite being essential for faithful chart reconstruction and consistent visual restyling. Several MLLM-based systems incorporate design-related capabilities as part of broader chart understanding and generation tasks. ChartLlama~\cite{han_chartllama_2023} instruction-tunes an MLLM for chart understanding, description, chart-to-code conversion, and editing. ChartReformer~\cite{yan_chartreformer_2024} enables natural-language-driven chart editing, ChartMimic~\cite{yang_chartmimic_2025} evaluates chart-to-code conversion as a proxy for cross-modal reasoning, and OneChart~\cite{chen_onechart_2024} extracts structural information using auxiliary tokens. While these systems demonstrate the broad capabilities of MLLMs for chart-related tasks, none systematically evaluate how reliably design properties, such as colors, marker shapes, and spatial configurations, can be extracted from arbitrary chart images.

In contrast to prior work, we first provide a systematic benchmark of five commercial MLLMs on both data and design extraction across five basic chart types. Building on these findings, we present a mixed-initiative system that combines MLLM-based extraction with human verification to support reliable and efficient chart reconstruction.
\section{Formative Study}

To understand the challenges and requirements of chart reconstruction from static chart images, we conducted a formative study with three visualization designers who regularly reconstruct charts from static chart images for external data communication. Each participant took part in a 30-minute semi-structured interview, and all sessions were documented through detailed interviewer notes.

\subsection{Current Challenges in Chart Reconstruction}

Participants reported recent use of MLLMs for chart reconstruction but identified several recurring challenges.

\textbf{\circled{C1} Data Accuracy and Verification Burden.}
Although MLLMs can extract data tables from chart images, designers emphasized that the extracted results still require manual verification before they can be trusted. P1 noted that ``\textit{we must manually validate data accuracy before using outputs---a critical requirement, especially in high-stakes contexts where visualizations inform medical and business decisions.}'' Likewise, P2 pointed out that ``\textit{generated charts may contain incorrect or mismatched data points, and errors in labels or headings.}''

\textbf{\circled{C2} Limited Output Editability.}
Participants also highlighted that current MLLM outputs are difficult to edit and refine. Rather than producing structured chart representations, existing systems often generate flat images that cannot be conveniently modified. As P2 explained, ``\textit{the output was hard to edit as a chart because they are non-editable (flat images) and lack layering (a single object instead of components).}'' Designers therefore require layered, component-based output formats that support subsequent editing and reuse.

\textbf{\circled{C3} Iterative, Multi-Step Workflows.}
Chart reconstruction is inherently iterative. Participants described repeatedly refining prompts and correcting outputs before obtaining satisfactory results. P3 explained that ``\textit{the workflow is iterative rather than one-shot, and multiple prompts and refinements are often needed.}''

\subsection{Design Goals}

Based on these findings, we derived three design goals.

\textbf{\circledVarD{G1} Accurate and Verifiable Output.}
The extracted data and design specifications should be accurate and easy to verify against the original chart. The system should support efficient side-by-side validation, allowing designers to identify and correct extraction errors with minimal manual effort.

\textbf{\circledVarD{G2} Editable Output.}
The reconstructed chart should remain fully editable as a structured visualization rather than a flat image. This enables designers to refine both visual properties and underlying data after reconstruction. To support such editing, the system should generate layered, component-based representations such as Vega-Lite specifications.

\textbf{\circledVarD{G3} Efficient Chart Reconstruction.}
The overall reconstruction workflow should be substantially more efficient than manual reconstruction, which typically requires 2--4 hours. At the same time, the system should minimize repetitive prompting and manual intervention while preserving opportunities for verification and iterative refinement.

\section{Benchmark Study}

To select an appropriate backbone model for \sys, we conducted a systematic evaluation of five commercial MLLMs across five basic chart types, assessing their performance on both data and design extraction tasks.

\subsection{Evaluation Setup}
\textbf{Target Models.}
We evaluated five commercially available MLLMs: \texttt{Claude Opus 4.6}~\cite{claude_4.6}, \texttt{Claude Opus 4.7}~\cite{claude_4.7}, \texttt{GPT-5.4}~\cite{gpt5.4}, \texttt{GPT-4o}~\cite{got-4o}, and \texttt{Nova Pro}~\cite{nova_pro}.

\textbf{Target Chart Types.}
Based on a survey of charts collected from scientific publications and publicly available company presentations, we selected five representative chart types for evaluation: bar charts, line charts, scatter plots, pie charts, and box plots. Together, these chart types account for approximately 78\% (1881/2421) of the charts in our corpus.

\textbf{Dataset Construction Pipeline.}
To enable systematic evaluation with ground-truth labels, we synthesized a dataset of 3,250 charts using Plotly\footnote{https://plotly.com/graphing-libraries/}. This approach is preferable to manually annotating real-world charts because it avoids prohibitive annotation costs while providing precise control over chart properties and accurate ground-truth annotations.

The dataset construction pipeline consisted of three stages:
1) \textit{Content Pool Creation:}
Seed content, including chart titles, axis labels, and representative data ranges, was extracted from real-world charts to ensure domain realism.
2) \textit{Synthetic Chart Generation:}
Charts were generated by systematically varying the configurable properties for each chart type using Plotly. Since each chart type has a different number of configurable properties, this process produced different numbers of charts per type.
3) \textit{Manual Verification and Sampling:}
All generated charts were manually inspected to identify and remove anomalies. The verified dataset was uniformly sampled across chart types, resulting in 450--900 charts per type: bar chart (900), scatter plot (400), line chart (850), pie chart (450), and box plot (650).


\textbf{Target Design Properties \& Evaluation Metrics.}
We identified a set of visually observable and commonly used design properties from Plotly's chart specification that are relevant to chart reconstruction.
Representative design properties include:

\begin{itemize}[itemsep=2pt]
    \item \textbf{Visual Attributes:}
    Position (e.g., axis tick positions in pixel coordinates), size (e.g., marker size and bar width), color (e.g., bar colors), and shape (e.g., marker symbols).

    \item \textbf{Structure and Metadata:}
    Structure (e.g., chart type and bar orientation), existence (e.g., the presence of error bars), and text (e.g., chart titles and legend labels).
\end{itemize}

Evaluation metrics were selected according to the data type of each property, as summarized in \cref{tab:evaluation-metrics}.

\begin{table}[h]
  \centering
  \small
  \begin{tabularx}{\columnwidth}{lll}
    \toprule
    \textbf{Property Type} & \textbf{Data Type} & \textbf{Evaluation Metric} \\
    \midrule
    Position & Numeric & Relative Error (\%) \\
    Size & Numeric & Relative Error (\%) \\
    Color & Color & Color Distance ($\Delta E 2000$) \\
    Shape & Categorical & Exact Match Error (\%) \\
    Structure & Categorical & Exact Match Error (\%) \\
    Existence & Categorical & Exact Match Error (\%) \\
    Text & Text & Character Error Rate (CER) (\%) \\
    \bottomrule
  \end{tabularx}
  \caption{Evaluation metrics for design property extraction.}
  \label{tab:evaluation-metrics}
\end{table}

\textbf{Target Data Fields \& Evaluation Metrics.}
For chart data extraction, we evaluated three aspects: textual data, numerical data, and output format correctness.

Text fields were evaluated using Mean Character Error Rate (MCER), which measures the average character-level difference between extracted and ground-truth text. For each chart, given predicted strings $\hat{s}_i$ and ground-truth strings $s_i$ over $n$ labels, let $\mathrm{LD}(\hat{s}_i,s_i)$ denote the Levenshtein distance and $|s_i|$ the length of the ground-truth string. MCER is defined as:

\begin{equation}
\mathrm{MCER}
=
\frac{1}{n}
\sum_{i=1}^{n}
\frac{\mathrm{LD}(\hat{s}_i,s_i)}{|s_i|}
\times100\%.
\end{equation}
Lower MCER indicates better text extraction accuracy.

Numerical values were evaluated using Adaptive MAPE (Mean Absolute Percentage Error)~\cite{he_making_2026}, which measures the average absolute relative error across extracted numerical values. Let $V_{\max}$ denote the maximum absolute ground-truth value. Adaptive MAPE is defined as:

\begin{equation}
\mathrm{Adaptive\ MAPE}
=
\frac{1}{n}
\sum_{i=1}^{n}
\left|
\frac{\hat{v}_i-v_i}{V_{\max}}
\right|
\times100\%.
\end{equation}
Lower Adaptive MAPE indicates higher numerical accuracy.

Finally, we evaluated format error rate by comparing the extracted table structure with the expected number of rows and columns. When the extracted format was incorrect (e.g., missing or additional data points), the extracted values could not be aligned with the ground truth. In such cases, both MCER and Adaptive MAPE were set to 100\% to indicate complete extraction failure.

\subsection{Results}

\begin{figure}[ht]
    \centering
    \includegraphics[width=\columnwidth]{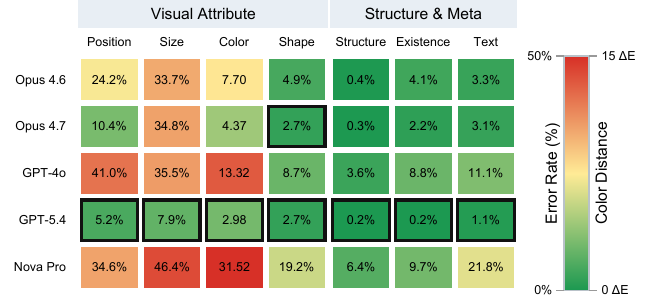}
    \caption{Evaluation results of design extraction across different property categories.}
    \label{fig:design-extraction-results}
\end{figure}

\textbf{Design Extraction.}
\texttt{GPT-5.4} consistently achieved the best performance across all design property categories among the five evaluated models (see \cref{fig:design-extraction-results}). Key findings include:
\begin{itemize}[nosep]
    \item Categorical and text properties (shape, structure, existence, and text) were reliably extracted by all models except \texttt{Nova Pro}, with error rates below 11.1\%.
    \item Numeric properties (position and size) and color remained substantially more challenging across all models.
    \item Only \texttt{GPT-5.4} maintained consistently low error rates across all property types, including numeric and color.
\end{itemize}
\noindent Overall, these results suggest that current MLLMs can reliably recover categorical and textual design properties but still struggle with quantitative visual attributes that require precise visual grounding.

\begin{figure}[t]
    \centering
    \includegraphics[width=\columnwidth]{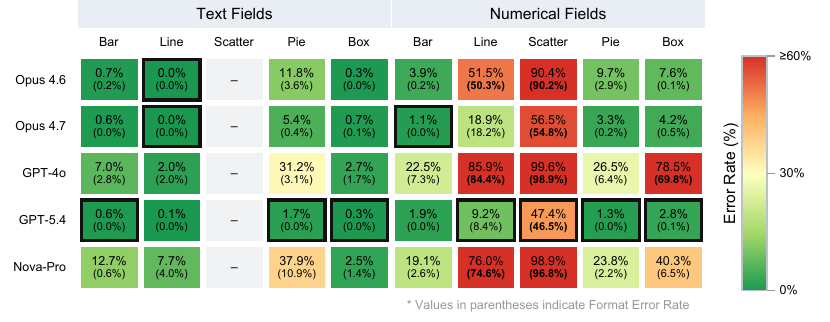}
    \caption{Evaluation results of data extraction across different chart types.}
    \label{fig:data-extraction-results}
\end{figure}

\textbf{Data Extraction.}
Text data extraction was consistently accurate across all evaluated models, while numerical data extraction exhibited substantial variation across chart types (see \cref{fig:data-extraction-results}). Key findings include:
\begin{itemize}[nosep]
    \item Text data was reliably extracted across all models with low error rates.
    \item Numerical data exhibited substantial variation: bar charts and pie charts achieved lower Adaptive MAPE, whereas scatter plots (47.4\%--99.6\%) and line charts (9.2\%--85.9\%) remained considerably more challenging.
    \item Format errors, namely extracting an incorrect number of data points, were a major source of failure, particularly for line charts with dashed or dotted styles and dense scatter plots.
\end{itemize}
\noindent Overall, numerical extraction remains the primary bottleneck for current MLLMs, especially for chart types requiring precise spatial reasoning.

Based on these findings, we selected \texttt{GPT-5.4} for \sys because it consistently achieved the best overall performance across both data and design extraction tasks, particularly on numeric and spatial extraction where other models exhibited significant weaknesses.
\section{System Design and Implementation}

\sys is a web-based mixed-initiative system designed to fulfill \circledVarD{G1}-\circledVarD{G3}.

\subsection{System Architecture}\label{sec:architecture}

The system follows a four-stage pipeline: 
1) \textit{Chart Type Detection:}
The uploaded chart image is classified into one of the five chart types using the \texttt{GPT-5.4}.
2) \textit{Specification Extraction:}
\texttt{GPT-5.4} extracts both the data table and design specifications.
The extracted design properties vary by chart type (e.g., bar width and error bars for bar charts, marker shape and size for scatter plots).
3) \textit{Chart Reconstruction:}
A Plotly chart is generated from the extracted data and design specifications.
Plotly was chosen because it supports scientific chart elements (error bars, statistical annotations), and generates output compatible with JavaScript, Python, and R.
4) \textit{User Verification and Refinement:}
Users can verify and refine the extracted data and edit the reconstructed chart design using the interactive interface. The interface breaks down into four panels in the editor mode and four comparison options in the comparison mode, as described in the next subsection.

\subsection{MLLM-Driven Extraction}

The extraction pipeline (stages 1--2 in \cref{sec:architecture}) uses \texttt{GPT-5.4} with chart-type-specific prompts and programmatic post-processing to recover specifications from static chart images.

\textbf{Chart-Type-Specific Prompt Composition.}
The extraction process consists of two sequential MLLM calls: the first classifies the chart type, and the second extracts the full specification (data table and design properties).
Rather than using a monolithic prompt, the system composes extraction prompts from reusable blocks: a shared block for common properties (e.g., title, legend, background color) and chart-type-specific blocks for unique properties (e.g., bar width for bar charts, marker shape for scatter plots).
Each prompt embeds a typed JSON schema defining the expected output structure, with inline behavioral notes providing disambiguation rules (e.g., log-scale value conversion and series ordering conventions).

\textbf{Post-Processing and Auto-Correction.}
The system applies programmatic post-processing to address known MLLM failure modes.
For example, when the model returns flattened data for grouped bar charts instead of the expected multi-series structure, the system detects repeated category labels and automatically reshapes the data into the correct format.
Additionally, when extracted data points fall outside the reported axis range, the system automatically expands the axis bounds to accommodate all values.

\subsection{User Interface Features}

To support efficient and quick reconstruction, we design a set of interactive features.

\textbf{Overlay-Based Verification (\circledVarD{G1}, \circledVarD{G3}):}
Extracted data points are rendered in a Data Table panel (\cref{fig:teaser}-a) and as colored markers overlaid on the original chart image in a Data Verification panel (\cref{fig:teaser}-b).
The two views are bidirectionally linked, allowing users to identify the correspondence between each table entry and its rendered point.
Users first calibrate the overlay markers by aligning the axes (\cref{fig:teaser}-e), then drag markers to correct positions; corresponding data table values update in real time.

\textbf{Chart Edit (\circledVarD{G2}, \circledVarD{G3}):}
The reconstructed chart is displayed in the Reconstructed Chart panel (\cref{fig:teaser}-d) and linked with the Data Table panel (\cref{fig:teaser}-a) and the Design panel (\cref{fig:teaser}-c), which support editing properties such as color, size, and axis labels.
Changes in either panel update the chart in real time.
To support rapid editing without iterative prompts, we employ GUI-based direct manipulation rather than chat-based LLM interaction.

\textbf{Chart Comparison (\circledVarD{G2}):}
Clicking the comparison button at the top-left corner opens the comparison view  (\cref{fig:teaser}-f) with four modes: \textit{Horizontal Split} (\cref{fig:teaser}-f1) and \textit{Vertical Split} (\cref{fig:teaser}-f2) for side-by-side alignment, \textit{Fade} (\cref{fig:teaser}-f3) for crossfading between the original and reconstructed charts via opacity adjustment, and \textit{Highlight} (\cref{fig:teaser}-f4) which renders the original in green and the reconstructed in red with overlapping regions in gray.






\section{Discussion}

Our evaluation and system have several limitations that suggest directions for future work.

\textbf{User Study Evaluation.}
Although we have received positive feedback from visualization designers during system demonstrations, a formal user study has not yet been conducted.
User studies with visualization designers are needed to validate the three design goals and assess usability and efficiency relative to manual reconstruction.

\textbf{Synthetic Data Limitations.}
Our benchmark relies on synthetic charts to enable systematic evaluation with ground-truth annotations.
Although the charts were generated using content derived from real-world examples, they do not fully capture the visual artifacts, compression effects, and stylistic diversity found in published charts.
Future work should evaluate our system on real-world chart images to better understand generalization performance.

\textbf{Chart Type Coverage.}
Our evaluation focuses on five basic chart types. More specialized visualizations, such as heatmaps and Kaplan--Meier (KM) curves~\cite{rich_practical_2010}, as well as composite figures, remain outside the current scope.
KM curves, step-function plots used in biomedical research to estimate survival probabilities, pose extraction challenges due to their unique geometry and censoring markers.
Heatmaps encode data through color scales rather than position and size, requiring dedicated decoding functions.
The proposed framework is extensible to these additional visualizations. 

\textbf{Model Adaptation for Chart Extraction.}
Our benchmark revealed that MLLMs are proficient at extracting categorical or text design properties and text data values but struggle with numerical design properties and quantitative data values.
Fine-tuning MLLMs or developing chart-specific adaptation methods using our synthetic dataset is a promising direction for improving extraction accuracy on these challenging cases.

\section{Conclusion}

We presented \sys, a mixed-initiative system for reconstructing charts from static chart images by extracting both the underlying data and visual design specifications. Through a systematic benchmark of five commercial MLLMs, we showed that current models reliably extract textual and categorical information but continue to struggle with numeric and spatial information, highlighting the need for interactive verification. Guided by these findings, \sys combines MLLM-based extraction with an overlay-based verification interface that supports real-time chart rebuilding and direct editing of extracted data and design properties. Our work highlights the potential of combining automatic extraction with human verification to support reliable and efficient chart reconstruction.

\bibliographystyle{abbrv-doi}

\bibliography{template}

@misc{web_plot_digitizer,
	title = {WebPlotDigitizer},
  howpublished={\url{https://automeris.io/}},
	url = {https://automeris.io/},
	language = {en},
	urldate = {2026-03-22},
	journal = {Google for Developers}
}

@inproceedings{masson_chartdetective_2023,
	address = {Hamburg Germany},
	title = {{ChartDetective}: {Easy} and {Accurate} {Interactive} {Data} {Extraction} from {Complex} {Vector} {Charts}},
	isbn = {978-1-4503-9421-5},
	doi = {10.1145/3544548.3581113},
	language = {en},
	urldate = {2026-05-21},
	booktitle = {Proceedings of the 2023 {CHI} {Conference} on {Human} {Factors} in {Computing} {Systems}},
	publisher = {ACM},
	author = {Masson, Damien and Malacria, Sylvain and Vogel, Daniel and Lank, Edward and Casiez, Géry},
	month = apr,
	year = {2023},
	pages = {1--17}
}

@inproceedings{he_making_2026,
	address = {Barcelona Spain},
	title = {Making {Multimodal} {LLMs} {Reliable} {Chart} {Data} {Extractors}: {A} {Benchmark} and {Training} {Framework}},
	isbn = {979-8-4007-2278-3},
	doi = {10.1145/3772318.3790721},
	language = {en},
	urldate = {2026-05-21},
	booktitle = {Proceedings of the 2026 {CHI} {Conference} on {Human} {Factors} in {Computing} {Systems}},
	publisher = {ACM},
	author = {He, Yuchen and Ying, Peizhi and Cheng, Liqi and Peng, Kuilin and Tian, Yuan and Deng, Dazhen and Wu, Yingcai},
	month = apr,
	year = {2026},
	pages = {1--18}
}

@incollection{yan_chartreformer_2024,
	address = {Cham},
	title = {{ChartReformer}: {Natural} {Language}-{Driven} {Chart} {Image} {Editing}},
	volume = {14804},
	isbn = {978-3-031-70532-8 978-3-031-70533-5},
	doi = {10.1007/978-3-031-70533-5_26},
	language = {en},
	urldate = {2026-05-26},
	booktitle = {Document {Analysis} and {Recognition} - {ICDAR} 2024},
	publisher = {Springer Nature Switzerland},
	author = {Yan, Pengyu and Bhosale, Mahesh and Lal, Jay and Adhikari, Bikhyat and Doermann, David},
	editor = {Barney Smith, Elisa H. and Liwicki, Marcus and Peng, Liangrui},
	year = {2024},
	pages = {453--469}
}

@misc{han_chartllama_2023,
	title = {{ChartLlama}: {A} {Multimodal} {LLM} for {Chart} {Understanding} and {Generation}},
	doi = {10.48550/arXiv.2311.16483},
	language = {en},
	urldate = {2026-05-29},
	publisher = {arXiv},
	author = {Han, Yucheng and Zhang, Chi and Chen, Xin and Yang, Xu and Wang, Zhibin and Yu, Gang and Fu, Bin and Zhang, Hanwang},
	month = nov,
	year = {2023}
}

@misc{liu_deplot_2023,
	title = {{DePlot}: {One}-shot visual language reasoning by plot-to-table translation},
	doi = {10.48550/arXiv.2212.10505},
	language = {en},
	urldate = {2026-06-01},
	publisher = {arXiv},
	author = {Liu, Fangyu and Eisenschlos, Julian Martin and Piccinno, Francesco and Krichene, Syrine and Pang, Chenxi and Lee, Kenton and Joshi, Mandar and Chen, Wenhu and Collier, Nigel and Altun, Yasemin},
	month = may,
	year = {2023}
}

@misc{chen_onechart_2024,
	title = {{OneChart}: {Purify} the {Chart} {Structural} {Extraction} via {One} {Auxiliary} {Token}},
	doi = {10.48550/arXiv.2404.09987},
  language = {en},
	urldate = {2026-07-05},
	publisher = {arXiv},
	author = {Chen, Jinyue and Kong, Lingyu and Wei, Haoran and Liu, Chenglong and Ge, Zheng and Zhao, Liang and Sun, Jianjian and Han, Chunrui and Zhang, Xiangyu},
	month = apr,
	year = {2024}
}

@misc{yang_chartmimic_2025,
	title = {{ChartMimic}: {Evaluating} {LMM}'s {Cross}-{Modal} {Reasoning} {Capability} via {Chart}-to-{Code} {Generation}},
	doi = {10.48550/arXiv.2406.09961},
  language = {en},
	urldate = {2026-07-22},
	publisher = {arXiv},
	author = {Yang, Cheng and Shi, Chufan and Liu, Yaxin and Shui, Bo and Wang, Junjie and Jing, Mohan and Xu, Linran and Zhu, Xinyu and Li, Siheng and Zhang, Yuxiang and Liu, Gongye and Nie, Xiaomei and Cai, Deng and Yang, Yujiu},
	month = feb,
	year = {2025}
}

@ARTICLE{huan_pixels_2025,
  author={Huang, Kung-Hsiang and Chan, Hou Pong and Fung, May and Qiu, Haoyi and Zhou, Mingyang and Joty, Shafiq and Chang, Shih-Fu and Ji, Heng},
  journal={IEEE Transactions on Knowledge and Data Engineering}, 
  title={From Pixels to Insights: A Survey on Automatic Chart Understanding in the Era of Large Foundation Models}, 
  year={2025},
  volume={37},
  number={5},
  pages={2550-2568},
  doi={10.1109/TKDE.2024.3513320}
}

@INPROCEEDINGS{islam_overview_2019,
  author={Islam, Mohaiminul and Jin, Shangzhu},
  booktitle={2019 International Conference on Information Science and Communications Technologies (ICISCT)}, 
  title={An Overview of Data Visualization}, 
  year={2019},
  volume={},
  number={},
  pages={1-7},
  doi={10.1109/ICISCT47635.2019.9012031}
}

@ARTICLE{chai_crowdchart_2021,
  author={Chai, Chengliang and Li, Guoliang and Fan, Ju and Luo, Yuyu},
  journal={IEEE Transactions on Knowledge and Data Engineering}, 
  title={CrowdChart: Crowdsourced Data Extraction From Visualization Charts}, 
  year={2021},
  volume={33},
  number={11},
  pages={3537-3549},
  doi={10.1109/TKDE.2020.2972543}
}

@INPROCEEDINGS{luo_chartocr_2021,
  author={Luo, Junyu and Li, Zekun and Wang, Jinpeng and Lin, Chin-Yew},
  booktitle={2021 IEEE Winter Conference on Applications of Computer Vision (WACV)}, 
  title={ChartOCR: Data Extraction from Charts Images via a Deep Hybrid Framework}, 
  year={2021},
  volume={},
  number={},
  pages={1916-1924},
  doi={10.1109/WACV48630.2021.00196}
}

@misc{nova_pro,
  title = {Amazon {{Nova}} Foundation Models},
  urldate = {2026-08-07},
  howpublished = {https://aws.amazon.com/nova/}
}

@misc{claude_4.6,
  title = {Introducing {{Claude Opus}} 4.6},
  urldate = {2026-08-07},
  howpublished = {https://www.anthropic.com/news/claude-opus-4-6}
}

@misc{got-4o,
  title = {Hello {{GPT-4o}}},
  urldate = {2026-08-07},
  howpublished = {https://openai.com/index/hello-gpt-4o/}
}

@misc{claude_4.7,
  title = {Introducing {{Claude Opus}} 4.7},
  urldate = {2026-08-07},
  howpublished = {https://www.anthropic.com/news/claude-opus-4-7}
}

@misc{gpt5.4,
  title = {Introducing {{GPT-5}}.4},
  urldate = {2026-08-07},
  howpublished = {https://openai.com/index/introducing-gpt-5-4/}
}

@article{rich_practical_2010,
  title={A practical guide to understanding Kaplan-Meier curves},
  author={Rich, Jason T and Neely, J Gail and Paniello, Randal C and Voelker, Courtney CJ and Nussenbaum, Brian and Wang, Eric W},
  journal={Otolaryngology—Head and Neck Surgery},
  volume={143},
  number={3},
  pages={331--336},
  year={2010},
  publisher={SAGE Publications Sage CA: Los Angeles, CA}
}
\end{document}